\documentclass[prb, twocolumn,tightenlines]{revtex4}
\usepackage{amsmath,amssymb,bm}
\usepackage{graphicx}
\usepackage{epstopdf}
\usepackage{latexsym}
\usepackage{adjustbox}
\usepackage{float}

\usepackage[lofdepth,lotdepth]{subfig}
\usepackage[justification=raggedright]{caption}

\usepackage{dsfont}	

\usepackage{color}
\usepackage{natbib}
\usepackage{braket}
\usepackage{hyperref}
\usepackage{comment}
\usepackage{array} 
\hypersetup{
  colorlinks,
  citecolor=magenta,
  linkcolor=blue,
  urlcolor=blue}
      
\DeclareMathAlphabet{\mathbbold}{U}{bbold}{m}{n}

\begin{document}

\title{First-order superfluid to valence bond solid phase transitions
  in easy-plane SU($N$) magnets for small-$N$}

\author{Jonathan D'Emidio}
\affiliation{Department of Physics \& Astronomy, University of
  Kentucky, Lexington, KY 40506-0055}

\author{Ribhu K. Kaul}
\affiliation{Department of Physics \& Astronomy, University of Kentucky, Lexington, KY 40506-0055}

\begin{abstract}
We consider the easy-plane limit of bipartite SU($N$) Heisenberg Hamiltonians which have a fundamental representation on one
sublattice and the conjugate to fundamental on the other
sublattice. For $N=2$ the easy plane limit of the SU(2) Heisenberg
model is the well known quantum XY model of a lattice superfluid. We introduce a logical method to
generalize the quantum XY model to arbitrary $N$, which
keeps the Hamiltonian sign-free. We show that these 
quantum Hamiltonians have a world-line representation as the statistical mechanics
of certain tightly packed loop models of $N$-colors in which neighboring loops are
disallowed from having the same color. In this loop representation we
design an efficient Monte Carlo cluster algorithm for our model. We
present extensive numerical results for these models on the two
dimensional square lattice, where we find the nearest neighbor model has
superfluid order for $N\leq 5$ and valence-bond order for $N> 5$. By
introducing SU($N$) easy-plane symmetric four-spin couplings we are able to tune across
the superfluid-VBS phase boundary for all $N\leq 5$. We present clear
evidence that this quantum phase
transition is first order for $N=2$ and $N=5$, suggesting that easy-plane
deconfined criticality runs away generically to a first order transition
for small-$N$.
\end{abstract}
\date{\today}
\maketitle

\section{Introduction}

Quantum phase transitions have emerged as an important paradigm in the
study of quantum many-body phenomena.~\cite{sachdev1999:qpt}
Deconfined criticality is a novel field theoretic proposal for a
continuous transition between a magnet or superfluid that breaks an
internal symmetry and a valence bond solid (VBS) that breaks a lattice
translational symmetry.~\cite{senthil2004:science,senthil2004:deconf_long} The field theories realized at these new
critical points are strongly coupled gauge theories, which
are rather fundamental and hence connected to a wide range of problems,~\cite{halperin1974:largeN,vishwanath2013:3dbti}
making the study of deconfined criticality of general interest in
theoretical physics. The study of deconfined critical points has been
significantly enhanced by the availability of sign problem free quantum
Monte Carlo (QMC) simulations which are able to access the strong
coupling physics of the emergent gauge theories in some particular
Marshall positive Hamiltonians that host this phase transition.~\cite{kaul2013:qmc}

Historically, in the discovery of deconfined criticality, a prominent role
was played by the ``easy-plane'' deconfined critical point, which is most naturally
realized in lattice models of superfluids.
The square lattice quantum XY model has been studied extensively in
the last few decades as the simplest quantum lattice spin model for a superfluid. The model can be written in the following ways,
\begin{eqnarray}
\label{eq:hxy1} H_{XY} &=& - J \sum_{\langle ij\rangle } \left ( S_i^x S_j^x +
   S_i^y S_j^y \right )\\
 &=& - \frac{J}{2} \sum_{\langle ij\rangle } \left ( S_i^+S_j^- +
   S_i^-S_j^+ \right ),
\end{eqnarray}
where the $\vec S_i$ are the usual $S=1/2$ Pauli matrices on site $i$. The
model has only a sub-group of the SU(2) symmetry of the Heisenberg
model: it has a $U(1)$ rotation symmetry about the $\hat z$ axis and a $Z_2$ symmetry of flipping the $S^z$ components. We
note here that the sign of $J$ is inconsequential, since it can be changed by a unitary
transformation. The
model $H_{XY}$ is Marshall positive and hence free of the sign problem
of quantum Monte  Carlo. Exploiting this fact, extensive numerical simulations have shown that $H_{XY}$ has long
range superfluid order at $T=0$. To study the destruction of superfluid order
in the ground state,
a sign-free generalization of the quantum XY model to include a four spin
coupling $K$ was introduced,~\cite{sandvik2002:hjk}
\begin{equation}
\label{eq:hk}
H_{K} = - K \sum_{\langle ijkl\rangle} \left (
  S_i^+S_j^-S_k^+S_l^-+S_i^-S_j^+S_k^-S_l^+ \right )
\end{equation}
It has been shown that this model hosts three phases, a superfluid
state at small $K/J$, valence-bond solid order for intermediate $K/J$ and
checkerboard ordered solid for large $K/J$. The phase transition between VBS
and checkerboard was found (as expected) to be strongly first order. On the other hand
the fate of the transition between superfluid and VBS in this model
has remained enigmatic: while no evidence for a discontinuity have been
observed in this model, large scaling violations may not be
interpreted consistently as a continuous
transition.~\cite{sandvik2006:unpub} This transition if continuous
would be the first studied example of a ``deconfined critical point,'' its first
numerical study~\cite{sandvik2002:hjk} even predating
the field theoretic proposal.~\cite{senthil2004:science} The square lattice easy-plane model
played a prominent role in the original deconfined critical point
proposal. Additionally it has been
shown that should the superfluid-VBS transition in the quantum XY model
 be continuous, it would be a rare example of a self-dual critical
 point,~\cite{motrunich2004:hhog} which can be connected to various interesting field theoretic
 formulations with topological terms.~\cite{senthil2006:topo} We note here that some direct studies of the effective
 field theory~\cite{kuklov2006:u1first,kragset2006:first} and other quantum models on
 frustrated lattices~\cite{isakov2006:kagome,sen2007:first} which are believed to be in the same
 universality class have found evidence for a first order
 transition. In the meantime, attention in deconfined criticality has
 shifted to the study of the fully SU($N$) symmetric
 Hamiltonians,~\cite{sandvik2007:deconf,lou2009:sun,kaul2012:j1j2} and
 more recently loop models,~\cite{nahum2015:deconf} which have shown compelling evidence for a
 continuous transition. 

However, given the important role that the square lattice XY model has had  as the first putative host of deconfined
criticality, it remains of interest to have an unambiguous answer
to the nature of the superfluid-VBS transition in this model. Instead
of addressing this issue in the J-K model, where the results of the simulations are hard to
interpret~\cite{sandvik2006:unpub} we study a different model which can be understood as an easy-plane generalization of
the SU(2) symmetric J-Q model.~\cite{sandvik2007:deconf} We provide clear evidence that the transition in this
model is direct and discontinuous both for superfluid and the VBS
order.  Additionally, we are able to
provide a simple way to
generalize the easy-plane J-Q model to larger-$N$. Through large scale numerical simulations
we show that $H^{N}_{J_\perp Q_\perp}$ hosts the superfluid-VBS phase transition
for all $N\leq 5$. We provide evidence that this transition remains first order
for $N=5$, which leads us to conclude that easy-plane deconfined
criticality is generically first order for small-$N$.

The paper is organized as follows. In Sec.~\ref{sec:ham}, we introduce
the new easy-plane J-Q model, $H^{N}_{J_\perp Q_\perp}$ and write it down in
different ways, emphasizing how it is connected to previously studied models. In Sec.~\ref{sec:loop} we show how it maps to a
particular family of loop models in one higher dimension and use this
representation to formulate an efficient Monte Carlo algorithm. In
Sec.~\ref{sec:sim}, we present results of simulations of
$H^{N}_{J_\perp Q_\perp}$ on the square lattice, focusing on the first-order nature of the phase
transitions in this model. Finally, in Sec.~\ref{sec:conc}, we provide
our conclusions and outlook.

\section{\label{sec:ham}Lattice Hamiltonians}

It is useful to write our Hamiltonian in a few different ways to
elucidate certain diverse aspects. First let us consider the
generalization of the XY model in the form, Eq.~(\ref{eq:hxy1}) to
arbitrary-$N$. A natural way to do this on a bipartite lattice in the
spirit of the work of Affleck,~\cite{affleck1985:lgN} is to consider the following Hamiltonian,
\begin{equation}
\label{eq:hjp1}
H^{N}_{J_\perp} =  -\frac{ J_\perp}{N} \sum_{\langle ij\rangle} \sum_{a \in od} T^a_i   
  T^{a*}_j,
\end{equation}
where  the $T_i$ are $N\times N$ matrices acting on site $i$, which are the generators of
SU($N$).  Here we choose the normalization such that $\mathrm{Tr}[T^aT^b]=\delta_{ab}$.
These generators act on a local Hilbert space which now can take $N$ different
colors and are denoted by $|\alpha\rangle_i$. We note that $i$ is always chosen on the A sub-lattice and $j$ is
chosen on the B sub-lattice, so that spins on one sublattice transform
in the fundamental and spins on the other sublattice transform under
the conjugate to fundamental representations. The important difference
with the usual fully SU($N$) symmetric Heisenberg model is
the sum on generators $a$, is restricted to the off-diagonal
generators. We note also that Eq. (\ref{eq:hxy1}) is equivalent to Eq. (\ref{eq:hjp1}) for $N=2$ up to a unitary transformation about $S^y$ on one sublattice.  It is well known that of the $N^2-1$ generators of SU($N$), $N-1$ are
diagonal and $N^2-N$ are off-diagonal. The restriction of the sum to
the $N^2-N$ off diagonal generators guarantees that when $N=2$, we
recover the usual XY model. For larger $N$, Eq.~(\ref{eq:hjp1}) is a convenient
large $N$ generalization of the XY model. We note here that the
restriction to off-diagonal generators reduces the symmetry from
SU($N$) to an Abelian U(1)$^{N-1}\times$S$_N$ subgroup (where S$_N$ is the permutation
group), which generalizes the U(1)$\times$Z$_2$ of
Eq.~(\ref{eq:hxy1}). Physically this corresponds to rotations about
the directions of the $N-1$ diagonal generators and a discrete
relabeling of the colors. We shall call this symmetry ep-SU($N$) symmetry.
Just like in the fully symmetric SU($N$) case, the virtue of staggering
the representation is that an A-B pair can form an
ep-SU($N$) singlet for all $N$.  For sites $i\in A$ and $j\in B$, the
singlet may be written as $ |s_{ij}\rangle = \frac{1}{\sqrt{N}} \sum_{\alpha=1}^N |\alpha_{i} \alpha_{j} \rangle $

Having written the model formally in terms of the SU($N$) generators helps us
see that model is a large-$N$ extension of $H_{XY}$. To get a
better intuition for the matrix elements of the model we now write it directly in terms of
the local Hilbert space. In terms of which, Eq.~(\ref{eq:hjp1}) takes
the simple form,
\begin{eqnarray}
\label{eq:hjp2} H^{N}_{J_\perp} &=&- \frac{ J_{\perp} }{N} \sum_{<ij>}\tilde{P}_{ij}\\
\tilde{P}_{ij}&=&\sum_{\alpha, \beta =1,\alpha \neq \beta}^N |\beta \beta\rangle_{ij}\langle \alpha \alpha |_{ij}.
\end{eqnarray}
We have put a tilde on $\tilde {P}_{ij}$ to emphasize that it is not a
projection operator. It would have been the SU($N$) symmetric projector on the singlet
state $|s\rangle_{ij}$ if the condition $\alpha\neq\beta$ were dropped. Having $\alpha
\neq \beta$ in $\tilde{P}_{ij}$ is a direct consequence of restricting
the sum on $a$ in Eq.~(\ref{eq:hjp1}) to off-diagonal generators and
is what results in the reduced easy-plane symmetry. This form of the
Hamiltonian also makes it apparent that the model is Marshall positive
(off-diagonal matrix elements are all negative) and is hence amenable
to sign problem free quantum Monte Carlo, just like the fully SU($N$)
symmetric case.~\cite{harada2003:sun}

Now it is straightforward to extend the idea of a four-spin coupling,
the so called ``Q'' interaction~\cite{sandvik2007:deconf,lou2009:sun} to the easy-plane case for any $N$,
\begin{equation}
H^{N}_{Q_\perp} =- \frac{Q_{\perp}}{N^2}\sum_{<ijkl>}\tilde{P}_{ij}\tilde{P}_{kl},
\end{equation}
where as in the original J-Q model, the sum is taken on both orientations of the bond pairing of the
square lattice plaquettes. We note here that for $N=2$, the $Q_\perp$
term does {\em not} reduce to $H_K$, Eq.~(\ref{eq:hk}). It includes
all the terms present there, but contains in addition pair hopping
terms that are not present in $H_K$. Thus $H^{N}_{J_\perp Q_\perp}\equiv
H^{N}_{J_\perp} + H^{N}_{Q_\perp}$ provides a new
way to cross the superfluid-VBS phase boundary for $N=2$ and also
provides a neat way to extend the four spin interaction to arbitrary
$N$, while still preserving the desired ep-SU($N$) symmetry.

\section{\label{sec:loop}Loop representation and Algorithm}

\begin{figure}[!t]
\centerline{\includegraphics[angle=0,width=0.4\columnwidth]{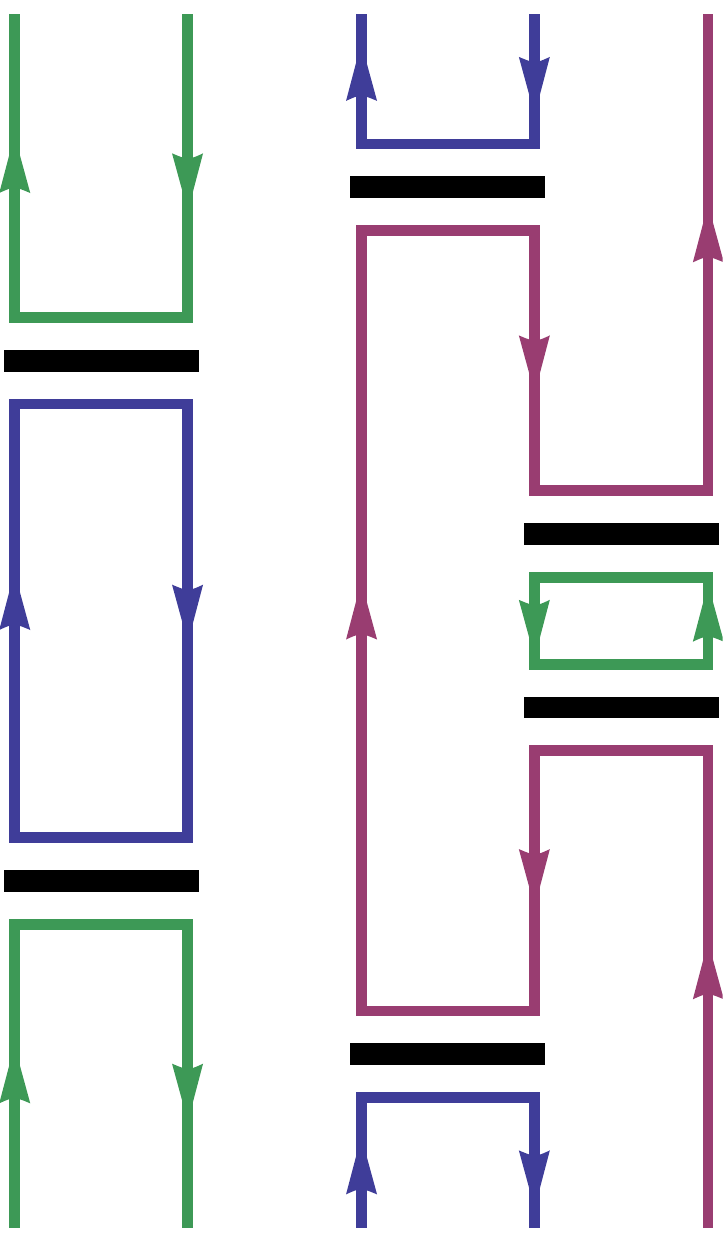}}
\caption{A section of a stochastic series expansion configuration of the Hamiltonian,
  Eq.~(\ref{eq:hjp2}), which takes a simple form in terms of $N$-colored tightly packed
  oriented closed loops.  By ``oriented'', we mean an orientation of
  the links (e.g. shown here as arrows going up on
  the A sub-lattice and down on the B sub-lattice) can be assigned
  before any loops are grown, and the orientation remains unviolated by
  each of the loops in a configuration.  The extreme easy-plane anisotropy results in
  an important constraint, {\em i.e.} loops that share a vertex (operators represented by the black bars) are restricted to have different colors.}
\label{fig:loops}
\end{figure}

It turns out that our models has a convenient world-line representation in terms
of closed  loops in one higher
dimension. While allowing us to design an efficient algorithm, it also
allows us to connect our work to recent numerical studies of
deconfined criticality in loop models.~\cite{nahum2015:deconf}

To see this start with
Eq.~(\ref{eq:hjp2}) and construct a stochastic series expansion (see Ref.~[\onlinecite{sandvik2010:vietri}] for a comprehensive review).
From previous work on similar models (see e.g. Ref.~[\onlinecite{kaul2014:design}]), we know that the partition
function of SU($N$) quantum spin Hamiltonians can be viewed as the
classical statistical mechanics of $N$-colors of tightly packed
oriented loops in one higher dimension (see also~\cite{nahum2011:loops}).  A
section of this particular kind of loop configuration, describing one term in the expansion of the partition function is depicted in FIG.~\ref{fig:loops} for our model.
An essential difference from the usual SU($N$) case (and the loop
models studied so far~\cite{nahum2011:loops,nahum2015:deconf}) is that loops which
are attached to the same vertex (operators represented by the black
bars) are restricted to have different colors, which is due to the
absence of diagonal matrix elements in
Eq.~(\ref{eq:hjp2}). This adds an important constraint: whereas in the
fully symmetric
SU($N$) case, given a legal loop configuration loop colors could be assigned
independently, in the easy-plane case studied  here only a
subset of the colorings (the ones that respect the constraint) are allowed. This
affects profoundly the physics of the easy-plane model, as well as the recoloring of loops in a Monte-Carlo algorithm.

\begin{figure}[!t]
\centering
\subfloat[Subfigure 1 list of figures text][Moves with probability 1/2]{
\includegraphics[width=0.225\textwidth]{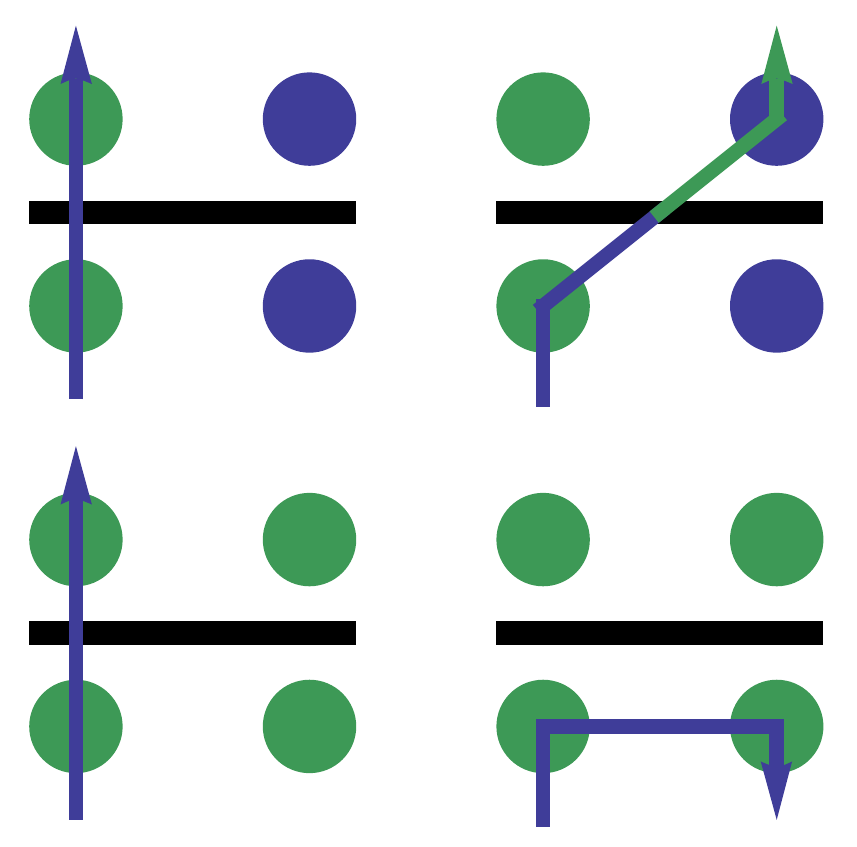}
\label{fig:subfig1}}
\qquad
\subfloat[Subfigure 2 list of figures text][Moves with probability 1]{
\includegraphics[width=0.1\textwidth]{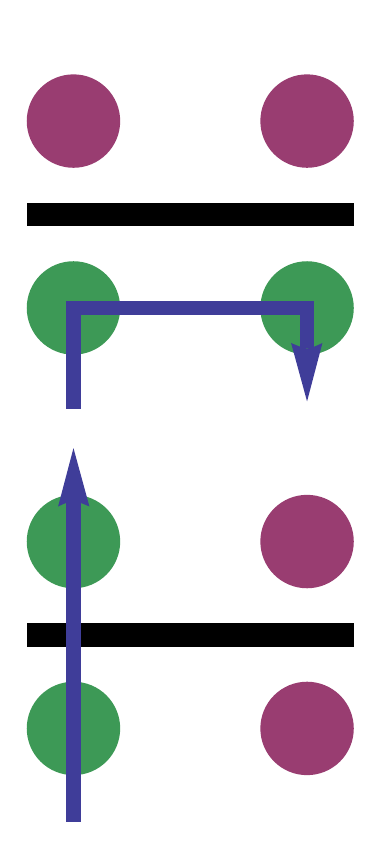}
\label{fig:subfig2}}
\caption{Loop updating moves describing how loops pass through a vertex, depicted with a black bar.  These moves cause conversions between the diagonal and off-diagonal matrix elements present in the Hamiltonian.  For each update pictured here there is a reverse process that we have not drawn.}
\label{fig:vtxupd}
\end{figure}

The fact that the loops in our model obey a coloring constraint and that diagonal operators are absent makes this model difficult to simulate as given.  This difficulty is overcome, as is commonly done, by shifting the Hamiltonian by a constant.  We shift our Hamiltonian in such a way that we introduce diagonal operators with equal weight as the off-diagonal ones.  This shift only adds a constant to the energies and does not affect the physics of the model in anyway, i.e. the eigenstates are identical.  The shifted Hamiltonian is given by
\begin{equation}
\label{eq:hjpShift}
\tilde{H}^{N}_{J_\perp} =- \frac{ J_{\perp} }{N} \sum_{<ij>}(\tilde{P}_{ij}+\mathds{1})
\end{equation}
where the diagonal and off-diagonal operators have equal weight.  With this shift the directed loop equations [\onlinecite{syljuasen2002:dirloop}], describing how the loops pass through a vertex, take on a particularly simple form.  We show the loop updating moves and corresponding probabilities in Fig. \ref{fig:vtxupd}.
With these rules the loops used to update our QMC configurations are not deterministic, as they are in the SU($N$) symmetric case.  The loops here can intersect with themselves and one may worry about growing a loop that fails to close onto itself.  We have never seen a case of a loop not closing during any of our simulations.

So far we have described our algorithm with $Q_{\perp}=0$.  The introduction of plaquette operators is easily dealt with if we regard them as simply a product of two (shifted) bond operators.  Our algorithm then inserts and removes diagonal plaquette operators, present due to the constant shift, and can convert them to off-diagonal plaquettes by performing loop updates.  Since the plaquette operator is viewed as two separate bonds operators, loop updates through the former are the same as for the latter.

\section{\label{sec:sim}Results of numerical simulations}

\begin{figure}[!t]
\centerline{\includegraphics[angle=0,width=1.0\columnwidth]{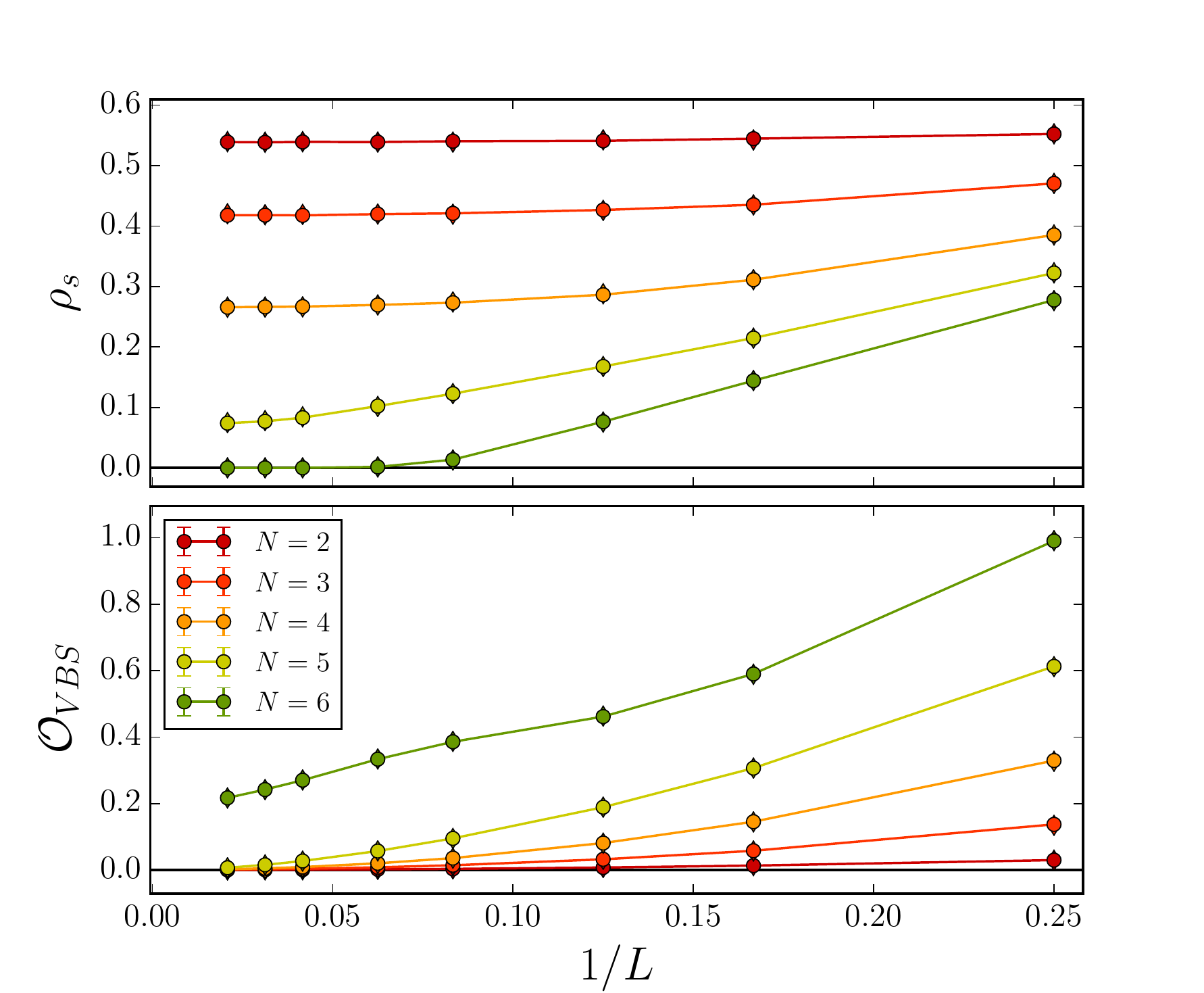}}
\caption{The stiffness and VBS order parameter extrapolations as a function of
  $1/L$ for different $N$ in the
  nearest neighbor easy-plane model $H^{N}_{J_\perp} $, defined in Eq.~(\ref{eq:hjp1})
  or equivalently~(\ref{eq:hjp2}). We find clear evidence that the
  system has superfluid order for all $N\leq 5$ and VBS order for
  $N>5$.  To show ground state convergence we plot both $\beta=L,2L$ in diamond and circular points, respectively.}
\label{fig:Jonly}
\end{figure}

\subsection{MEASUREMENTS}

Here we will outline the measurements that we use to characterize the phases of our model.  On the magnetic side of the transition the spin stiffness, or superfluid stiffness in the language of hard-core bosons, serves as a useful order parameter.  It is formally defined as follows:
{\allowdisplaybreaks
\begin{equation}
		\rho_{s}=\frac{1}{N_{\mathrm{site}}}\frac{\partial^2 \langle H(\phi) \rangle}{\partial \phi^2}\Bigr|_{\phi=0}
\label{eq:stiffED}
\end{equation}}
where $H(\phi)$ means that we twist the boundary conditions along either the x or y direction.  This twist can be implemented, for instance, by attaching phase factors to all x-oriented bond operators relative to one color e.g. $e^{i \phi} |\alpha \alpha\rangle_{ij}\langle \bigcirc \bigcirc |_{ij}$ for $\alpha \neq \bigcirc$, $i\in A$ sublattice and $j \in B$ sublattice.  The Hermitian conjugate of this operator appears with $e^{-i \phi}$.  Due to the staggered representation, the signs in the exponent are flipped when $i\in B$  and $j \in A$.  The bond operators take on this form when the spins are incrementally twisted along the x-direction about the $N-1^{\mathrm{th}}$ diagonal generator of SU($N$).  In practice, due to the permutation symmetry of the model, the $\bigcirc$ color is arbitrary, and we can average over the colors in QMC.

In QMC the stiffness is related to the winding number of configurations via
\begin{equation}
		\rho_{s}=\frac{1}{N_{\mathrm{site}}}\frac{\langle W^2 \rangle}{\beta},
\label{eq:stiffQMC}
\end{equation}
where operators with a positive (negative) phase factor count as positive (negative) winding.

\begin{figure}[t]
\centerline{\includegraphics[angle=0,width=1.0\columnwidth]{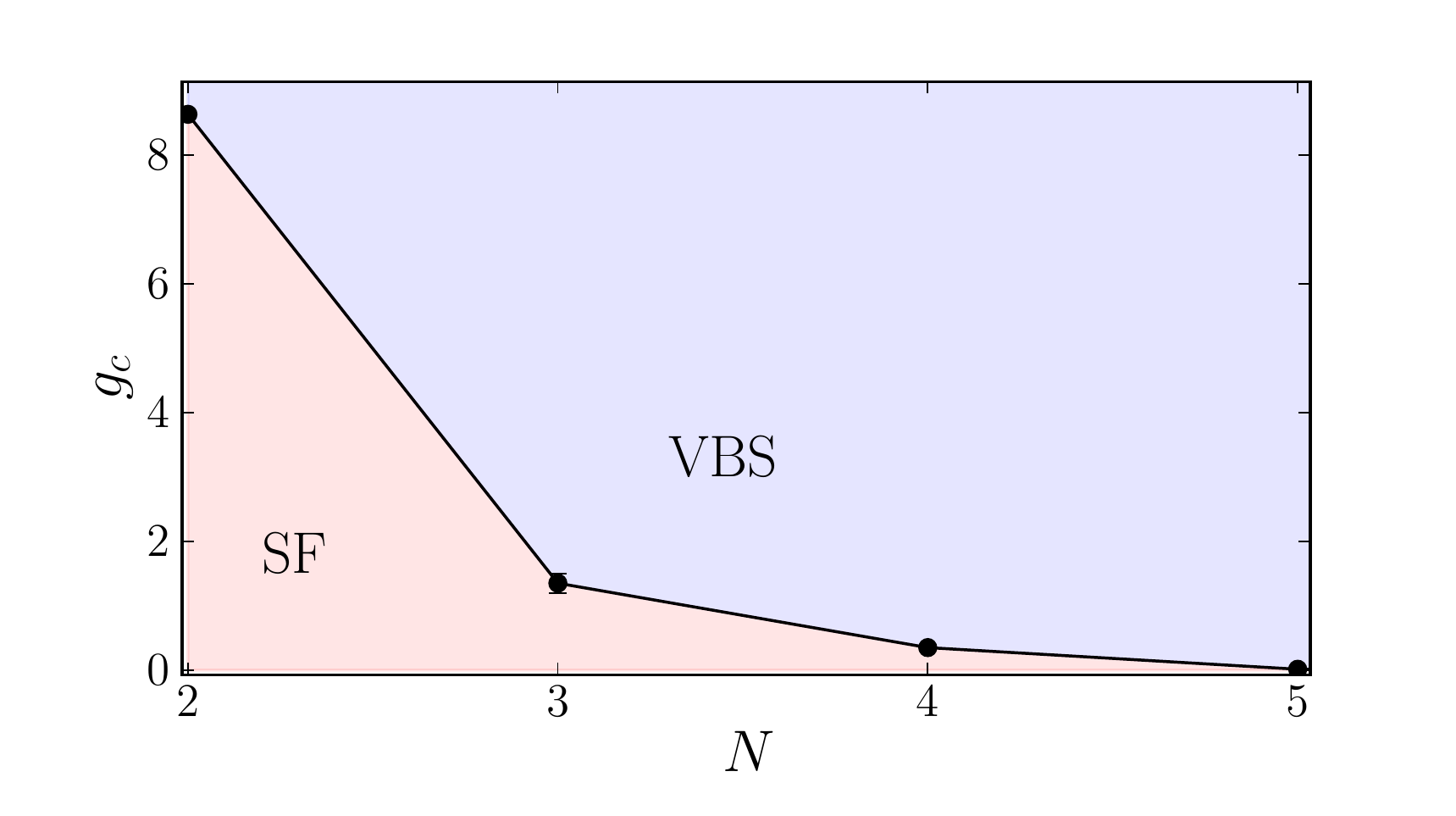}}
\caption{Phase diagram of the $H^{N}_{J_\perp Q_\perp}$ model as a
  function of the ratio $g_c=Q_\perp/J_\perp$ and $N$. Using the
  $H^{N}_{J_\perp Q_\perp}$ model we have access to the superfluid-VBS
phase boundary for $N=2,3,4$ and 5. In this work, we provide clear evidence that the
transition for $N=2$ and $N=5$ is direct but discontinuous for both
order parameters, suggesting that the easy-plane-SU($N$)
superfluid-VBS transition is generically first order for small-$N$.}
\label{fig:phasediag}
\end{figure}

In order to characterize the VBS phase, we measure the equal time
bond-bond correlation function $\langle
\tilde{P}_{\vec{r}\alpha}\tilde{P}_{\vec{r}'\alpha}\rangle$.   Here we have
denoted a bond by its location on the lattice $\vec{r}$ and its
orientation $\alpha$ ($x$ or $y$ in two-dimensions).  In the VBS phase, lattice translational symmetry is broken giving rise to a Bragg peak in the Fourier transform of the bond-bond correlator defined as
{\allowdisplaybreaks
\begin{equation}
		\tilde{C}^{\alpha}(\vec{q})=\frac{1}{N_{\mathrm{site}}^2}\sum_{\vec{r},\vec{r}'}e^{i(\vec{r}-\vec{r}')\cdot\vec{q}}\langle
                \tilde{P}_{\vec{r}\alpha}\tilde{P}_{\vec{r}'\alpha}\rangle.
\label{eq:FT}
\end{equation}}
For columnar VBS patterns, peaks appear at the momenta $(\pi,0)$ and $(0,\pi)$ for $x$ and $y$-oriented bonds, respectively.  The VBS order parameter is thus given by
{\allowdisplaybreaks
\begin{equation}
		\mathcal{O}_{VBS}=\frac{\tilde{C}^{x}(\pi,0)+\tilde{C}^{y}(0,\pi)}{2}.
\label{eq:Ovbs}
\end{equation}}

Another useful quantity that we use to locate phase transitions is the VBS ratio.
{\allowdisplaybreaks
\begin{equation}
		\mathcal{R}^x_{VBS}=1- \tilde{C}^{x}(\pi+2\pi/L,0)/\tilde{C}^{x}(\pi,0)
\label{eq:RXvbs}
\end{equation}}
And similarly for $\mathcal{R}^y_{VBS}$ with all of the $q_x$ and $q_y$ arguments swapped.  We then average over $x$ and $y$- orientations.
{\allowdisplaybreaks
\begin{equation}
		\mathcal{R}_{VBS}=\frac{\mathcal{R}^x_{VBS}+\mathcal{R}^y_{VBS}}{2}.
\label{eq:RXvbs}
\end{equation}}
This quantity goes to 1 in a phase with long-range VBS order, and
approaches 0 in a phase without VBS order.  It is thus a useful crossing quantity that allows us to locate the transition.

To construct these quantities, we measure the equal time bond-bond correlation function in QMC with the following estimator
{\allowdisplaybreaks
\begin{equation}
		\langle \Theta_1 \Theta_2\rangle=\frac{1}{\beta^2} \langle (n-1)! N[\Theta_1,\Theta_2] \rangle
\label{eq:QMCcorr}
\end{equation}}
where $\Theta_1$ and $\Theta_2$ are any two QMC operators (in our case off-diagonal bond operators), $n$ is the number of non-null operators in the operator string, and $N[\Theta_1,\Theta_2]$ is the number of times $\Theta_1$ and $\Theta_2$ appear in sequence in the operator string (excluding null slots).

\subsection{$J_\perp$-only model}
\begin{figure}[!t]
\centerline{\includegraphics[angle=0,width=1.0\columnwidth]{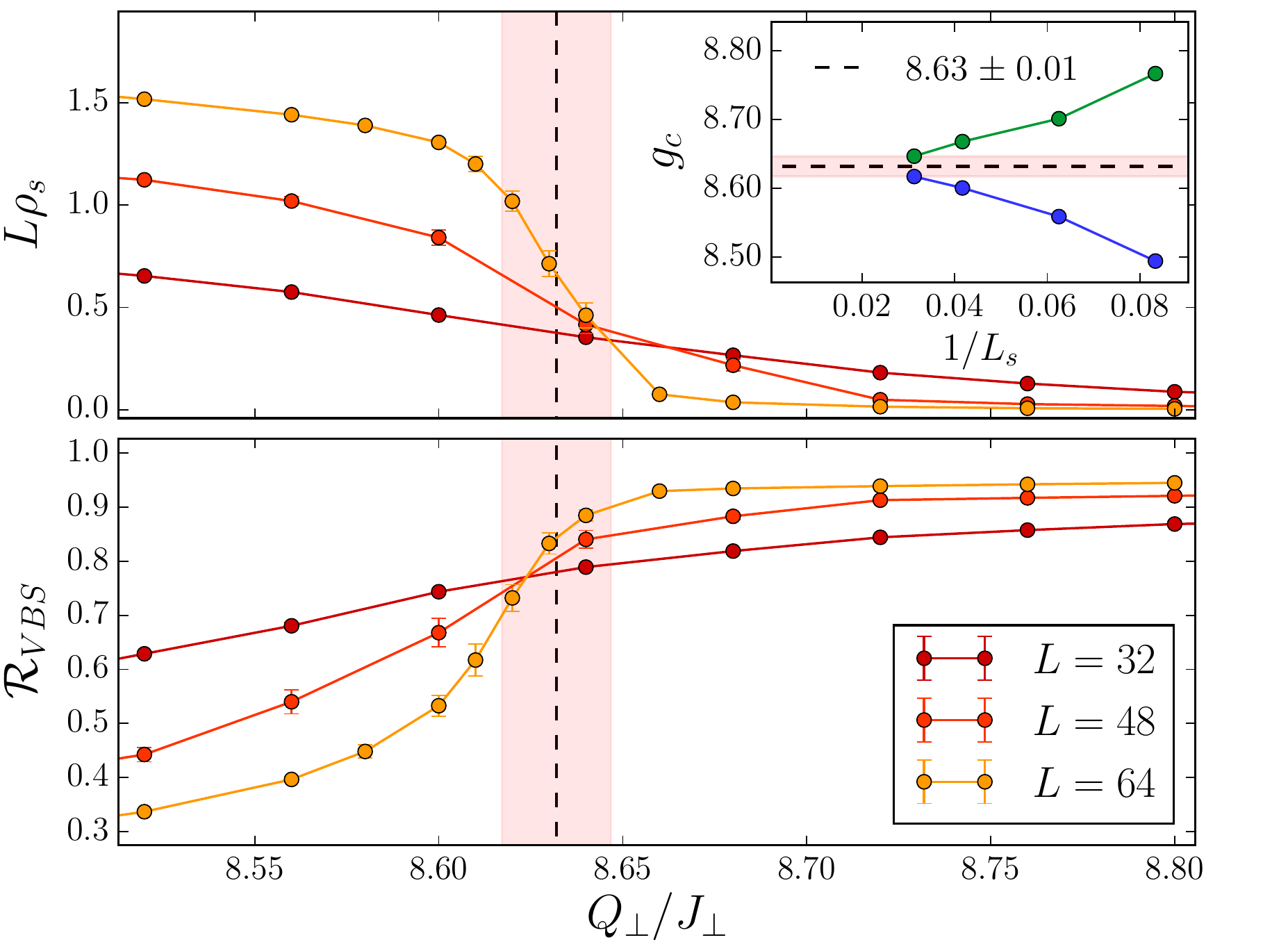}}
\caption{Crossings at the SF-VBS quantum phase transition for $N=2$ in
$H^{N}_{J_\perp Q_\perp}$. The main panels shows the crossings for
$L\rho_s$ and ${\cal R}_{VBS}$, which signal the destruction and onset
of SF and VBS order respectively. The inset shows that the SF-VBS
transition is direct, i.e. the destruction of SF order is accompanied
by the onset of VBS order at a coupling of $g_c=8.63(1)$.}
\label{fig:SU2cross}
\end{figure}

Consider the ground state phase of $H^{N}_{J_\perp} $ as we vary the number of colors $N$. For
$N=2$ we know the Hamiltonian is equivalent to the quantum XY model, whose
ground state is a superfluid. Here it is useful
to recall our
interpretation of $H^{N}_{J_\perp} $ as a classical loop model,
discussed in Sec.~\ref{sec:loop}. In the loop language this
corresponds to a long loop phase, where the loops span the system. It is expected
that when $N$ is increased the system would like to form as many loops
as possible to maximize its entropy, causing it to enter a short loop
phase. The way our loop model is defined this will cause the lattice
symmetry to break, leading to a VBS and destruction of superfluid order.   

Motivated by these considerations, in
FIG.~\ref{fig:Jonly} we plot the spin-stiffness and VBS order
parameter for $2\leq N \leq 6$.  We find that the ground state of
this Hamiltonian has long-range superfluid order for $N \leq 5$
(signaled by a finite spin stiffness) and has
VBS order for $N > 5$. These results are obtained by fixing
\(J_{\perp}=N, Q_{\perp}=0\) and we have plotted two values of inverse
temperature $\beta=L,2L$.  On the scale of the plot the two values of
$\beta$ are indistinguishable thus indicating ground state
convergence.  Since we find magnetic order for  $N \leq 5$, we can
study the phase transition for these values of $N$ with the
introduction of $Q_{\perp}$. 

In conclusion, we find that magnetic order
gives way to VBS order in the ep-SU($N$) magnet at an $N$ between 5
and 6. In comparison, in the
fully SU($N$) symmetric case magnetic order is lost between $N=4$ and 5.~\cite{harada2003:sun,beach2009:sun}

\subsection{$J_\perp$-$Q_\perp$  model}

\begin{figure}[!t]
\centerline{\includegraphics[angle=0,width=1.0\columnwidth]{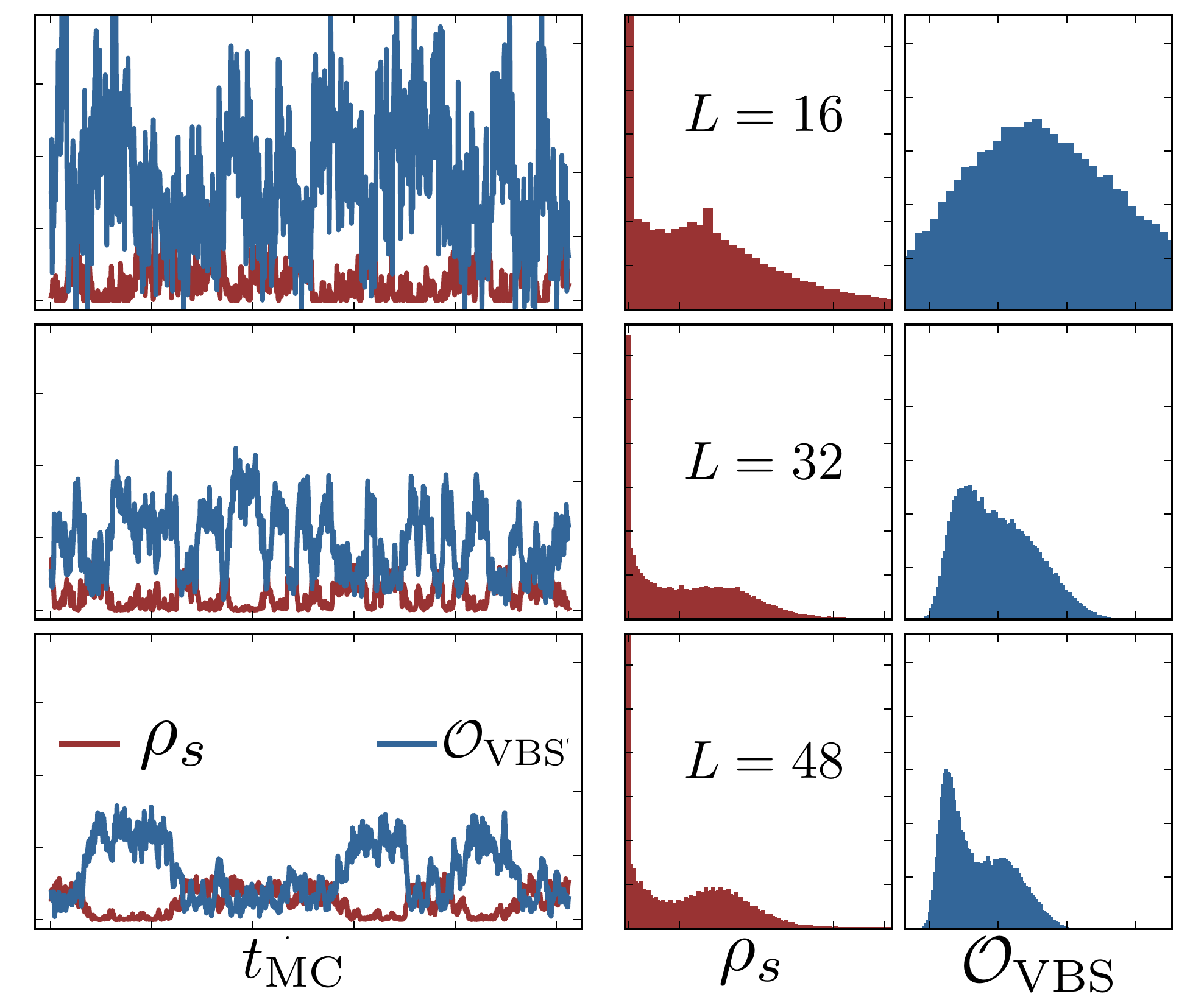}}
\caption{Evidence for first order behavior at the SF-VBS quantum phase
  transition for $N=2$. The data was collected at a coupling
  $g_c=8.63$.
The left panel shows MC histories for both $\rho_s$ and ${\cal O}_{VBS}$, with clear evidence for switching behavior characteristic of a first
  order transition. The right panel shows histograms of the same quantities
with double peaked structure which gets stronger with system
  size, again clearly indicating a first order transition.  The bin size for the histories is 400 MC 
steps per point}
\label{fig:SU2hist}
\end{figure}

Through numerical simulations described below we have obtained a phase diagram of the model $H^{N}_{J_\perp Q_\perp}$, which we show in
FIG.~\ref{fig:phasediag}. It is clear that the $Q_\perp$ interaction, which
mediates an attraction between the singlets will favor the formation
of a VBS state. We confirm this by numerical simulations in which we
find that the superfluid order is destroyed at a finite value of
$Q_\perp$ for $N=2$ and gives way to a VBS state. As we
increase $N$, the superfluid order becomes weaker in
$H^{N}_{J_\perp}$, hence requiring only a smaller value of the
$Q_\perp$ coupling to destroy the SF order, as is evident from the
phase diagram. A significant difference from the phase
diagram~\cite{sandvik2002:hjk} of the $J$-$K$ model, Eq.~(\ref{eq:hk}), is the
absence of the checkerboard ordered state at large $Q_\perp$, which is
present for large $K$. This can be expected on physical grounds by a
comparison of the $Q_\perp$ and $K$ terms. Finally, as expected for $N>5$, we are unable to
cross the superfluid-VBS phase boundary with $H^{N}_{J_\perp
  Q_\perp}$. 

We now turn to an analysis of the quantum phase
transitions which are denoted by the solid black points in FIG.~\ref{fig:phasediag}.
We begin with the case of $N=2$. FIG. \ref{fig:SU2cross} shows   SU(2)
data of our magnetic and VBS crossing quantities, $L \rho_s$ and
$\mathcal{R}_{VBS}$. In our data scans we have fixed \(J_{\perp}^2+ Q_{\perp}^2 = 1 \) and have set \( \beta=2 L \).  Our best estimate for the location of the transition is \(Q_{\perp} / J_{\perp}=8.63(1)\).  Close to the transition we find that signs of first-order behavior begin to appear.  FIG.~\ref{fig:SU2hist} shows the histories of our measurements as a function of Monte Carlo time.
We find clear signs that our measurements switch between two values close the transition, an effect which becomes more pronounced at larger system size.  Furthermore, binning this data into histograms shows a clear double peaked structure, indicating a first order transition (FIG.~\ref{fig:SU2hist}).

It is interesting to ask whether the first order behavior observed is
special to $N=2$. Given the self-duality of
$N=2$,~\cite{motrunich2004:hhog} which cannot be extended to $N>2$, it is possible that
the nature of the phase transition is different for other values of
$N$. To investigate this issue we have carried out a full numerical
study of $N=5$, the largest value of $N$ at which the SF-VBS
transition is accessible in our model, $H^{N}_{J_\perp Q_\perp}$. We find first
order behavior for $N=5$ very similar to what we have presented here
for $N=2$.  Numerical results and an analysis are presented in
Appendix~\ref{app:N5}. This leads us to conclude that the first-order
behavior is not special to $N=2$, but is present generically at
small-$N$ in our ep-SU($N$) models.

\section{\label{sec:conc}Conclusions}

To summarize, we have introduced a new family of ep-SU($N$) models
which generalize the $XY$ model of $N=2$. The generalization preserves
the property that it takes only two sites to form a singlet,
independent of the value of $N$. This property is important to the
formation of a columnar VBS state.
The generalization
defines an easy-plane four-spin interaction $Q_\perp$, which allows us
to access the superfluid-VBS phase boundary for $N=2,3,4$ and
5. Numerical studies show clear evidence for first order behavior. 

How
should this observation be interpreted in the terms of the field
theoretic picture of deconfined criticality?
In the deconfined
criticality scenario the first order transition could either be
because the field theory itself does not have a fixed point or because the
field theoretic fixed point is unstable to the introduction of quadrupled
monopoles (i.e. they are relevant at the fixed
point in the renormalization group sense).~\cite{block2013:fate} Given
the original arguments for irrelevance~\cite{senthil2004:deconf_long} and the numerical evidence in the symmetric case
that quadrupled monopoles are irrelevant, we
interpret the observed first order behavior here to imply that the
non-compact easy-plane deconfined field theory itself is
unstable to runaway flow for small-$N$. This is consistent with direct
numerical studies of the easy-plane theory for $N=2$.~\cite{kuklov2006:u1first,kragset2006:first} Our study raises questions that need to be addressed in future
field theoretic and numerical work. Is there a stable large-$N$
easy-plane SU($N$) deconfined critical point, or is there a generic reason it
does not exist? If it does exist, it is interesting to ask whether
numerical simulations will be able to access
a large enough $N$ to study this new quantum criticality.

Another interesting topic is the nature of the transition in
the ep-SU($N$) {\em without} Berry phases. This study can be
numerically achieved by studying our model, Eq.~(\ref{eq:hjp1}) on a bilayer square
lattice. For $N=2$ one would expect a 3D XY transition -- what happens
at larger-$N$? Previous work on the bilayer SU($N$)
magnet~\cite{kaul2012:bilayer}  and classical loop
models~\cite{nahum2013:loops} has shown in the symmetric case that the
transition is first order for large enough $N$, the answer to
the same question in the easy-plane case is not known currently and
will be pursued in future work.

{\em Acknowledgements:}
We acknowledge useful discussion with G.~Murthy. Partial financial support was received through NSF DMR-1056536.  The numerical simulations reported in the manuscript were carried out on the DLX cluster at the University of Kentucky.

\bibliographystyle{apsrev}
\bibliography{career}

\appendix

\section{QMC vs ED}

\begin{table}[h]
\centering
\begin{adjustbox}{max width=\textwidth}
\begin{tabular}{||l l l l l l||}
\noalign{\hrule height 2pt}
\multicolumn{1}{||c|}{$4 \times 4$} & \multicolumn{1}{c|}{$N=2$} & \multicolumn{1}{c|}{$J_{\perp}=1.0$} & \multicolumn{1}{c|}{$Q_{\perp}=0.0$} & \multicolumn{2}{c||}{$\beta_{\mathrm{qmc}}=48.0$}\\
\noalign{\hrule height 2pt}
\multicolumn{1}{||c|}{$e_{\mathrm{ex}}$} & \multicolumn{1}{l|}{$ -0.562486$} & \multicolumn{1}{c|}{$\rho_{\mathrm{ex}}$} & \multicolumn{1}{l|}{$0.27714$} & \multicolumn{1}{c|}{$\mathcal{O}_{\mathrm{ex}}$} & \multicolumn{1}{l||}{$0.007497$} \\
\multicolumn{1}{||c|}{$e_{\mathrm{qmc}}$} & \multicolumn{1}{l|}{$-0.562473(7)$} & \multicolumn{1}{c|}{$\rho_{\mathrm{qmc}}$} & \multicolumn{1}{l|}{$0.27710(3)$} & \multicolumn{1}{c|}{$\mathcal{O}_{\mathrm{qmc}}$} & \multicolumn{1}{l||}{$0.007497(1)$} \\
\noalign{\hrule height 2pt}
\multicolumn{1}{||c|}{$4 \times 4$} & \multicolumn{1}{c|}{$N=2$} & \multicolumn{1}{c|}{$J_{\perp}=1.0$} & \multicolumn{1}{c|}{$Q_{\perp}=1.0$} & \multicolumn{2}{c||}{$\beta_{\mathrm{qmc}}=32.0$}\\
\noalign{\hrule height 2pt}
\multicolumn{1}{||c|}{$e_{\mathrm{ex}}$} & \multicolumn{1}{l|}{$-0.741775$} & \multicolumn{1}{c|}{$\rho_{\mathrm{ex}}$} & \multicolumn{1}{l|}{$0.35858$} & \multicolumn{1}{c|}{$\mathcal{O}_{\mathrm{ex}}$} & \multicolumn{1}{l||}{$0.011097$} \\
\multicolumn{1}{||c|}{$e_{\mathrm{qmc}}$} & \multicolumn{1}{l|}{$-0.741770(8)$} & \multicolumn{1}{c|}{$\rho_{\mathrm{qmc}}$} & \multicolumn{1}{l|}{$0.35860(3)$} & \multicolumn{1}{c|}{$\mathcal{O}_{\mathrm{qmc}}$} & \multicolumn{1}{l||}{$0.011096(1)$} \\
\noalign{\hrule height 2pt}
\multicolumn{1}{||c|}{$4 \times 2$} & \multicolumn{1}{c|}{$N=3$} & \multicolumn{1}{c|}{$J_{\perp}=1.0$} & \multicolumn{1}{c|}{$Q_{\perp}=0.0$} & \multicolumn{2}{c||}{$\beta_{\mathrm{qmc}}=32.0$}\\
\noalign{\hrule height 2pt}
\multicolumn{1}{||c|}{$e_{\mathrm{ex}}$} & \multicolumn{1}{l|}{$-0.74157$} & \multicolumn{1}{c|}{$\rho^{x}_{\mathrm{ex}}$} & \multicolumn{1}{l|}{$0.043932$} & \multicolumn{1}{c|}{$\mathcal{O}^{x}_{\mathrm{ex}}$} & \multicolumn{1}{l||}{$ 0.030816$} \\
\multicolumn{1}{||c|}{$e_{\mathrm{qmc}}$} & \multicolumn{1}{l|}{$-0.74158(1)$} & \multicolumn{1}{c|}{$\rho^{x}_{\mathrm{qmc}}$} & \multicolumn{1}{l|}{$0.043928(5)$} & \multicolumn{1}{c|}{$\mathcal{O}^{x}_{\mathrm{qmc}}$} & \multicolumn{1}{l||}{$0.030821(3)$} \\
\noalign{\hrule height 2pt}
\multicolumn{1}{||c|}{$4 \times 2$} & \multicolumn{1}{c|}{$N=3$} & \multicolumn{1}{c|}{$J_{\perp}=1.0$} & \multicolumn{1}{c|}{$Q_{\perp}=1.0$} & \multicolumn{2}{c||}{$\beta_{\mathrm{qmc}}=32.0$}\\
\noalign{\hrule height 2pt}
\multicolumn{1}{||c|}{$e_{\mathrm{ex}}$} & \multicolumn{1}{l|}{$-1.25095$} & \multicolumn{1}{c|}{$\rho^{x}_{\mathrm{ex}}$} & \multicolumn{1}{l|}{$0.011196$} & \multicolumn{1}{c|}{$\mathcal{O}^{x}_{\mathrm{ex}}$} & \multicolumn{1}{l||}{$0.029878$} \\
\multicolumn{1}{||c|}{$e_{\mathrm{qmc}}$} & \multicolumn{1}{l|}{$-1.25095(2)$} & \multicolumn{1}{c|}{$\rho^{x}_{\mathrm{qmc}}$} & \multicolumn{1}{l|}{$0.011199(2)$} & \multicolumn{1}{c|}{$\mathcal{O}^{x}_{\mathrm{qmc}}$} & \multicolumn{1}{l||}{$0.029878(2)$} \\
\hline
\end{tabular}
\end{adjustbox}
\caption{Test comparisons of measurements from exact diagonalization and
 finite-$T$ QMC studies for the $N=2$ and $N=3$.  The energies reported here are per site and the stiffness
and VBS order parameters are defined in equations  (\ref{eq:stiffED} , \ref{eq:stiffQMC}) and (\ref{eq:Ovbs}).  For rectangular
systems we have used the stiffness along the $x$-direction and VBS order parameter for $x$-oriented bonds.  All systems have periodic
boundary conditions.}
\label{tab:qN}
\end{table}

For future reference, Table~\ref{tab:qN} contains test comparisons between measurements obtained from
a SSE-QMC study and exact diagonalization (ED) on $4\times 4$ and $4 \times
2$ systems with various $J_{\perp}$-$Q_{\perp}$ at $N=2$ and $N=3$.  We list values for the ground state
energy per site, spin stiffness as in Eqs. (\ref{eq:stiffED} , \ref{eq:stiffQMC}) as well as the VBS order parameter,
defined in Eq. (\ref{eq:Ovbs}).  For the rectangular systems we have indicated that we use the stiffness in the $x$-direction
and VBS order parameter with $x$-oriented bonds.

\section{\label{app:N5}$N=5$}

\begin{figure}[!h]
\centerline{\includegraphics[width=1.0\columnwidth]{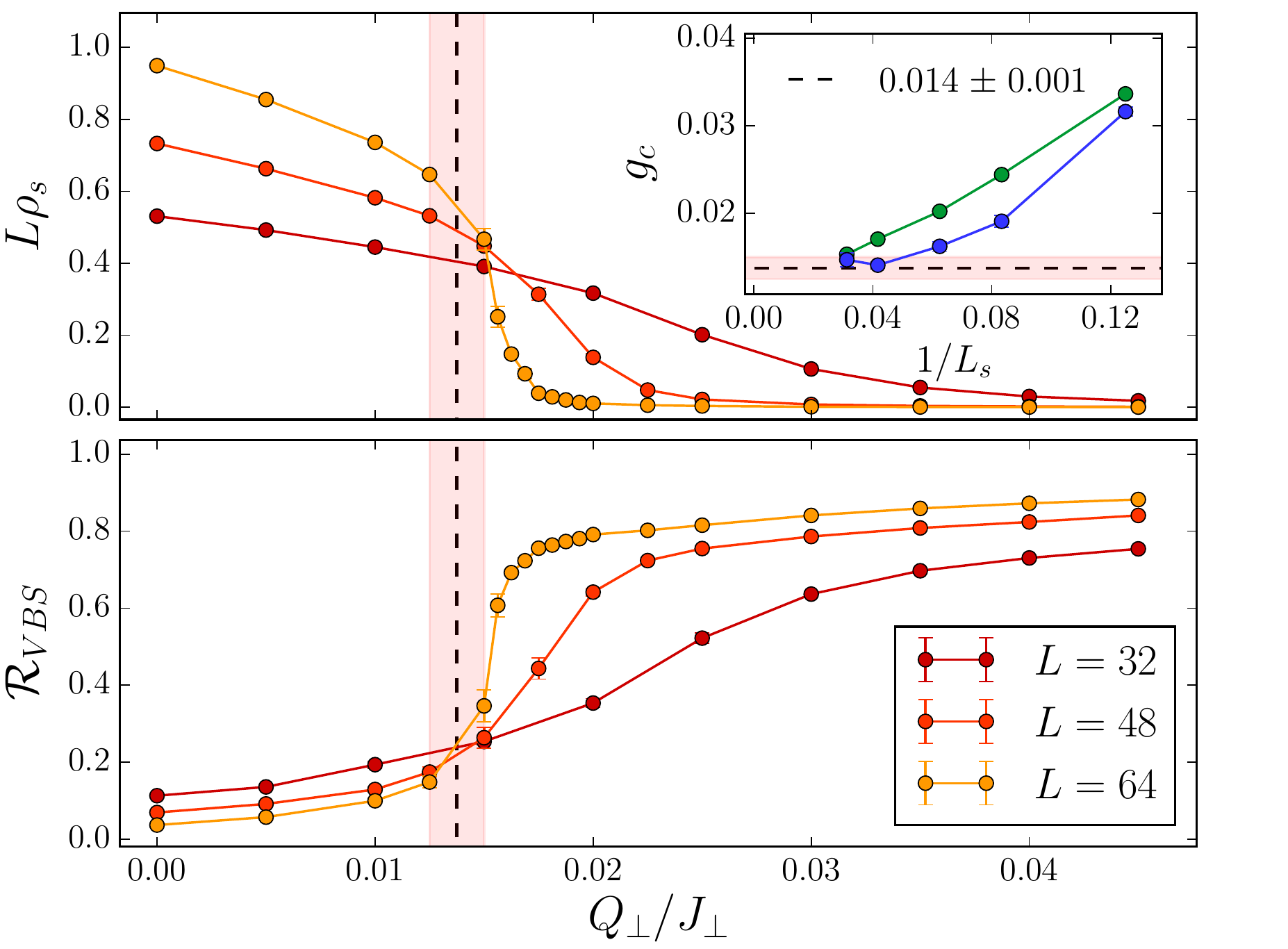}}
\caption{Crossings at the SF-VBS quantum phase transition for $N=5$ in
$H^{N}_{J_\perp Q_\perp}$. The main panels shows the crossings for
$L\rho_s$ and ${\cal R}_{VBS}$, which again show a direct transition between 
the superfluid and VBS states at $g_c=0.014(1)$.}
\label{fig:SU5cross}
\end{figure}

Here we present a study of the SF-VBS phase transition at $N=5$, which
also shows clear symptoms of first-order behavior.  In Fig. \ref{fig:SU5cross} 
we show the SF-VBS crossings that allow us to determine the critical
point.
Unlike in the $N=2$ case, the position of the crossings drifts in the same direction from both the SF and VBS side.
Taking finely binned data near the transition again shows us signs of
first-order behavior.  

\begin{figure}[H]
\centerline{\includegraphics[angle=0,width=1.0\columnwidth]{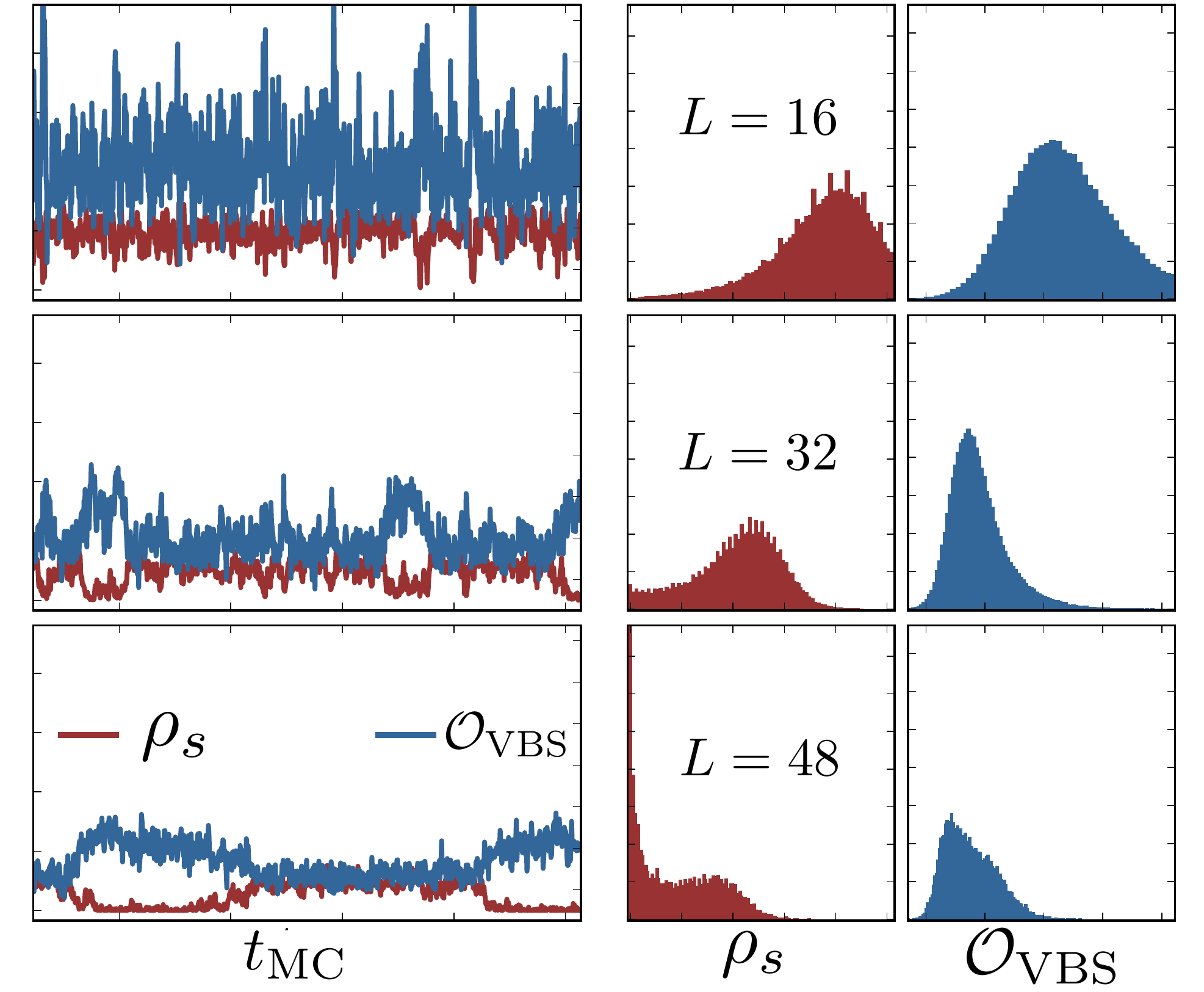}}
\caption{Evidence for first order behavior at the SF-VBS quantum phase
  transition for $N=5$. The data was collected at a coupling
  $g=0.01875$.
The left panel shows MC histories for both $\rho_s$ and ${\cal O}_{VBS}$, with clear evidence for switching behavior characteristic of a first
  order transition at the largest system size $L=48$. The right panel shows histograms of the same quantities
with double peaked structure emerging at $L=48$.  Here we note that the location of the transition for each system size drifts more significantly than in the $N=2$ case.  Also it can be argued that the transition shows signs of weakening.  Here we use a finer bin size for the histories (100 MC steps per point).}
\label{fig:SU5hist}
\end{figure}

In Fig. \ref{fig:SU5hist} we show history and histogram data for $N=5$.
For our largest system size we are able to see significant evidence of a first order transition in both the history and histogram.  We note here that unlike in the $N=2$ case, the location of the transition seems to drift substantially with the system size.  This results in only seeing clear signs of on-off switching (indicative of a first-order transition) for our largest system size, although it serves to illustrate the presence of a drift.  It is arguably the case that the transition is weakening as a function of $N$ given the less pronounced double peaks in the histograms.  

\end{document}